# Fractal Properties of Multiagent News Diffusion Model

**D.V. Lande, V.A. Dodonov**
Institute for Information Recording of NAS of Ukraine, Kiev

*The paper deals with fractal characteristics (Hurst exponent) and wavelet-scaleograms of the information distribution model, suggested by the authors. The authors have studied the effect of Hurst exponent change depending upon the model parameters, which have semantic meaning. The paper also considers fractal characteristics of real information streams. It is described, how the Hurst exponent dynamics depends on these information streams state in practice*

### Introduction

Statistical studies of time-series, corresponding to the volume of informative messages streams on some or other topic in the global networks are of great importance in analyzing substantial processes, which are reflected in these networks. However, along with studying common statistic properties of time-series, wavelet-analysis and fractal analysis has been recently used with increased frequency for solving forecasting problems, revealing periodicities, anomalies.

This paper will describe a model of topic-based news diffusion. In order to formally define topic-based information streams, let us make some assumptions, common for all further discussion [1].

### Information stream

Let us consider an interval $(a,\tau)$ of a real axis (time axis), where $\tau > a$. Let us assume, that in this time interval in accordance with particular patterns (one of the possible patterns will be suggested later on) some quantity of information documents are published in the network – $k$. On the time axis we will designate the moments, when separate documents are published, by $\tau_1, \tau_2, ..., \tau_k$ $(a \leq \tau_1 \leq \tau_2 \leq ... \leq \tau_k \leq \tau)$. Let us denote, that information stream is a process $N_\alpha(\tau)$, the implementation of which is characterized by a number of points (documents), which appeared in the interval $(a,\tau)$, as a function of the right end of the interval $\tau$. Thus, the implementation of information stream is a non-decreasing staircase-like, always integer-valued function $N_\alpha(\tau)$.

This definition is true to fact in local time domains, but it does not take into consideration such effect, as ageing of information, which is inconsistent with "accumulating" property of information stream $N_\alpha(\tau)$ in large time spans, which is taken into account in the model, proposed below.

In a narrower sense, by topic-based information stream we shall mean a quantity of documents, which in a certain sense (in content) match the specified topic. In the following paragraphs we shall consider a general picture of topic-based information streams dynamics.

Different approaches are used to model information streams. Among these approaches one can name nonlinear analytical models, models based upon the cellular automata concept [2-3], multiagent models [4-5].

The complexity of operation processes of informative messages' interrelation system, their origin and organization of impact upon the society makes it necessary to study the respective mechanisms and, consequently, to develop models or entire simulation complexes.

In the following paragraphs we shall turn our attention to multiagent concept of information diffusion. As distinguished from the well-known approaches, where people's behavior is simulated, in this case the simulation object is the informational space, which is considered as a model environment of information agents. Interconnected substances – informative messages are regarded as these agents. Precisely informative messages are considered as an instrument of information influence.

**Multiagent model**

Let us consider a multiagent model with the following performance parameters. Informative messages can be replicated (by way of "reposting"), they can contain links both to informative messages of similar content and to other objects of the real and the virtual world, they can "die" due to ageing etc. [4]. The agent's evolution will be connected with the events, which happen to such agent. As regards the principal characteristic, let us introduce the "energy" ($E$), which reflects the timeliness of the message and the degree of interest to it. It goes without saying, that ageing of information or negative reaction will reduce the message's energy, and positive reaction or appearance of the link to such message will increase its energy.

The agent appears with the initial energy $E_0$ and with each discrete time marking its energy decreases by 1. We shall consider events, which are typical for social networks: "like", "repost", "link" (providing the link reference to one agent by another agent). These events make an impact upon the agent's energy in the following way: "like" increases the energy by 1, "repost" increases the energy by 2, "link" increases the energy by 1. On the other hand, the probability, that one of these events will take place, depends upon the timeliness of the message, the interest to the information contained in it, which in terms of the model is expressed by energy. In this regard let us estimate the probability of a certain event having happened to the message with energy $E$ in the following way:

$$p_{like}^{(E)} = p_{l_0}\varphi(E); \quad p_{repost}^{(E)} = p_{r_0}\varphi(E),$$

where $p_{l_0}$, $p_{r_0}$ are model parameters, and $\varphi$ is some monotonic non-decreasing function of the agent's current energy with values in [0, 1]. When the energy drops down to 0, the agent "dies".

Modeling the dynamics of the whole information stream starts from one agent. A new agent can appear by two ways. The first way consists in copying the existing agent with the "repost" operation. The agent's self-generation is also possible, which corresponds to publishing a new message. Thus at any specific time with particular probabilities any of the events can happen to each of the agents. Also at any specific time with a probability $p_s$ a new agent can appear due to self-generation.

Let us consider a life journey of one agent. The agent appears with an initial energy value $E_0$ and further on its energy changes depending upon the events, which happen to him. Les us suppose, that two events are possible: "like" and "repost". One of these events, both of them simultaneously or none of them may happen per unit time.

Let us designate by $\varepsilon_t$ the agent's energy value at the time moment $t$. Then the energy value at the next time moment can be written down in the following way

$$\varepsilon_{t+1} = \varepsilon_t + \delta_t,$$

where $\delta_t$ is a random variable with values in {-1, 0, 1, 2}. According to the rules of energy change, described above, an increase in the energy by 2 corresponds to simultaneous "like" and "repost"; an increase by 1 – to "repost" only; the energy remains the same, if there is only "like"; and decreases by 1, if none of the events has happened.

The process of the agent's energy change can be regarded as integer-valued random walk. Since the energy value at the next time moment depends only upon the energy value at the previous time moment, the stochastic sequence $(\varepsilon_0, \varepsilon_1, ..., \varepsilon_t, ...)$ in the initial form is a Markovian chain with some transition probabilities $p_{ij}$.

In contrast with the paper [4], where "rigid" parameters of the model $p_{l_0}$, $p_{r_0}$ are set, in this paper, in particular, a "soft" model has been studied, which changes the aforementioned parameters thereof during functioning.

As time-series for the research, we consider the population volume of agents-messages, functioning in the multiagent environment in each particular moment of time. The results of modeling, as well as of wavelet-analysis [6-8] and diagram of Hurst exponent [9-11] for such time series in each particular case are given below.

**Wavelet-analysis of information streams**

Wavelet-analysis is built upon wavelet transform, which represents a special type of linear transformation, the basic functions of which (wavelets) have specific properties. Wavelet (small wave) is some function, concentrated in the small neighborhood and abruptly declining to zero with distance from it both in time domain and frequency domain. There are various wavelets, marked by different properties. At the same time, all wavelets have the form of short wavepackets with zero integral value, localized on the time axis, which are shift-independent and scale-independent.

Two operations can be applied to any wavelet:
– shift, that is moving its localization are in time;
– scaling (extension or compression).

The main idea of wavelet transform is that nonstationary time series is divided into separate intervals (so-called "watch windows"), and in each of them one should calculate a scalar product (a value, which characterizes the degree of similarity of two patterns) of the data under study with different shifts of some wavelet in different scales. Wavelet transform generates a series of coefficients, with the help of which the initial series is represented. They are functions of two variables: time and frequency, and therefore they form a surface in three-dimensional space. These coefficients show, to what extent the process behavior at the given point is analogous to wavelet in this scale. The more similar is the form of the analyzed relation in the neighborhood of the given point to the form of wavelet, the higher absolute value is reached by the respective coefficient.

The technique of using wavelets makes it possible to reveal occasional and irregular "spikes", abrupt changes in quantitative parameters during different time periods, in particular, volume of topic-based publications in the web-space. In this case it is possible to reveal the moments of cycles' generation, as well as moments, when periods of regular dynamics are followed by chaotic oscillations.

The time series under discussion can be approximated by a curve, which, in its turn, can be represented by the sum of harmonic oscillations of different frequency and amplitude. Low-frequency oscillations are responsible for slow, smooth, large-scale changes of the initial series' values, and high-frequency oscillations – for short, small-scale changes. The more significant are the changes of the variable, described by this pattern in this scale, the greater amplitude is observed for the component of the respective frequency. Consequently, the time series under study can be considered in time-and-frequency domain – that is about studying the pattern, describing the process depending both upon time and frequency.

Continuous wavelet transform for a function $f(t)$ is built with the help of continuous scale transformations and transfers of the selected wavelet $\psi(t)$ with arbitrary values of the scale coefficient $a$ and the shift parameter $b$:

$$W(a,b) = (f(t), \psi(t)) = \frac{1}{\sqrt{a}} \int_{-\infty}^{\infty} f(t) \psi^* \left( \frac{t-b}{a} \right) dt.$$

The obtained coefficients are represented in graphic form as a chart of transformation coefficients, or a scaleogram. Along one axis of the scaleogram the wavelet shit is plotted (time axis), and along the other axis scales are plotted (axis of scales). After that, the points of the chart, which has been obtained, are colored depending upon the value of the respective coefficient (the higher is the coefficient, the brighter are the picture colors). A scaleogram makes visible all characteristic features of the initial series: scale and intensity of periodic changes, direction and value of trends, presence, layout and duration of local specifics.

As a case-study, let us consider the multiagent system, described above, with the following parameters: the probability of a new agent's appearance at each step makes up 0.9, the probability of "like" to the agent is equal to 0.05, and the probability of "reposting" from the agent is equal to 0.001. The top part of figure 1 represents the diagram of publications' volume (quantity of messages), present at each step of the system's evolution. The bottom part thereof shows the respective wavelet-scaleogram – the result of continuous wavelet-analysis (Gauss wavelet).

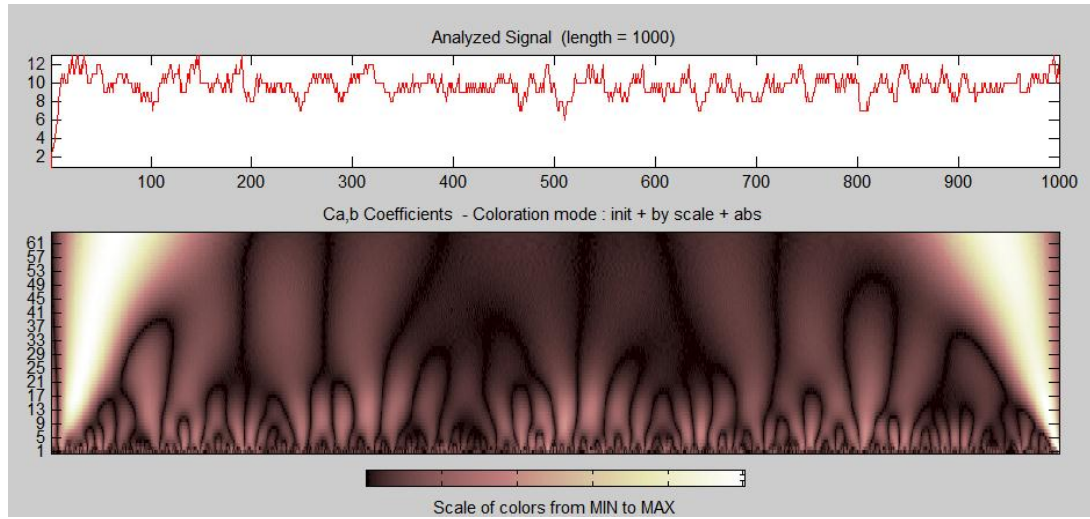

Fig. 1 – Case-study 1. Volume of agents' population and wavelet-scaleogram

Figure 2 shows a similar diagram for case-study 2, when the probability of "reposting" has increased to the value 0.05. In this case the time-series amplitude has augmented significantly, which is also evidenced by a lighter top part of the scaleogram. As we can see, both time-series are of random nature with similar elements at different amplitude levels.

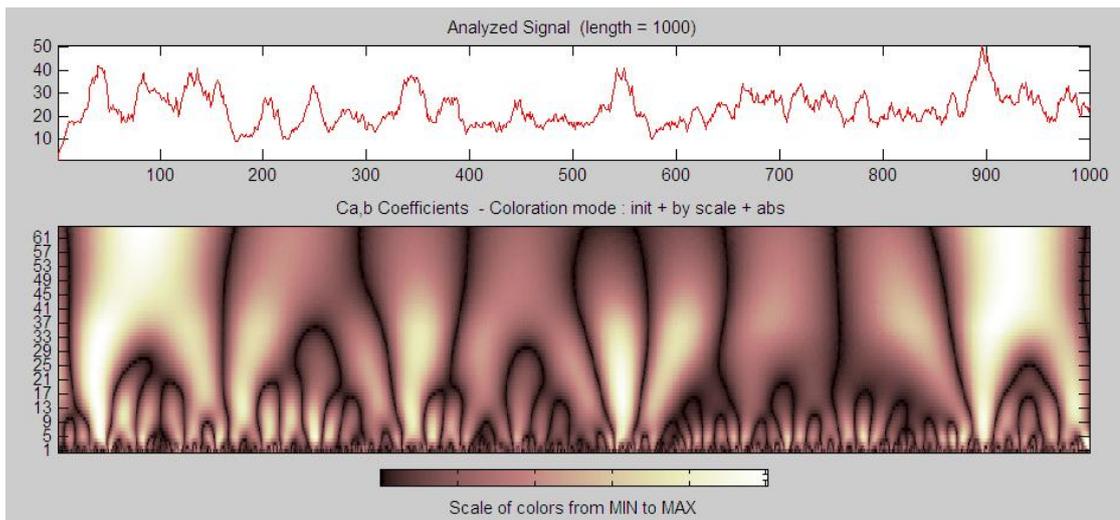

Fig. 2 – Case-study 2. Volume of agents' population and wavelet-scaleogram

The dynamics of real information streams often shows abrupt changes, related to changes in real processes, generating them. That is why, as case-study 3 we have studied the series, corresponding to the model, in which during the first 20% of evolution steps the "repost" parameter corresponded to case-study 1, and the remaining 80% – to case-study 2. The respective diagram and scaleogram are given in figure 3. As one should expect, the change of the process behavior can be clearly seen in the wavelet-scaleogram.

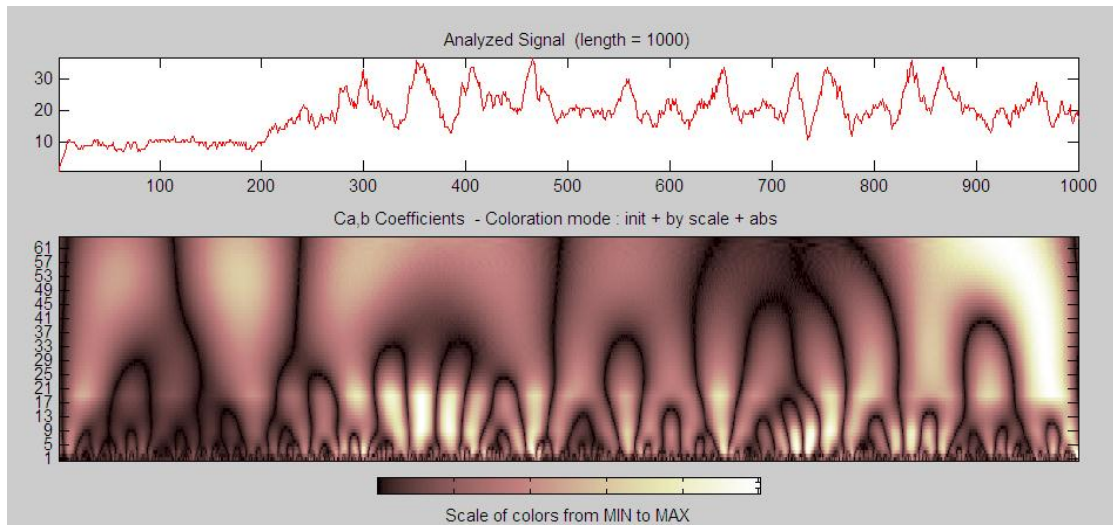

Fig. 3 – Case-study 3. Volume of agents' population and wavelet-scaleogram

As a case-study 4 let us consider a real information stream – the array of news items from the web-space, which arrived to the database of the content-monitoring system InfoStream (http://infostream.ua) on request "Tesla" (motor vehicle) during 2016. Figure 4 shows the dynamics of the respective messages' quantity and a wavelet-scaleogram, which in a sense proved to be similar to case-study 3. Figure 4 clearly shows the day of the year, when the process changed its behavior.

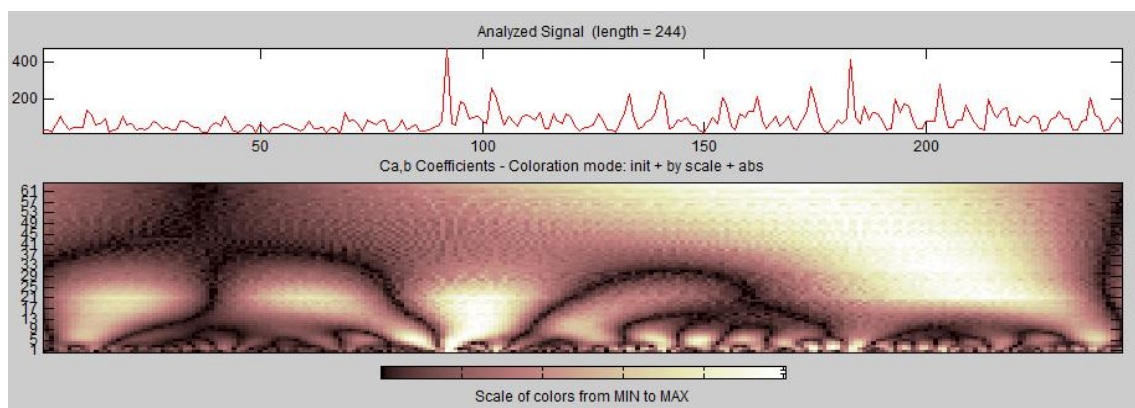

Fig. 4. Case-study 4. Dynamic of time-series, corresponding to the request "Tesla" and the respective wavelet-scaleogram.

This case-study shows that wavelet-analysis makes it possible to reveal not only evident anomalies in the series under study, but also critical values, which are hidden behind relatively small absolute values of the series elements. For example, the highest values are observed not only on the 90$^{th}$ day, when Tesla Motors automaker presented a budget-oriented version of its electric car, but also at unobvious extreme points, for example on the 178[th] day – the accident of a self-driving car Tesla Model S.

**Fractal analysis of information streams**

In order to investigate fractal characteristics of topic-based information streams we have studied the values of Hurst exponent (*H)* for the defined period for time series, composed of a number of relevant messages. Hurst exponent is related to the standardized range coefficient (*R/S*), where *R* is the "range" of the respective time-series, calculated in a particular way, and *S* – standard deviation. The dynamics of Hurst exponent was studied depending upon the model parameters. Hurst exponent for a time series is calculated in accordance with the following algorithm. At first the average value of the measured variable is calculated (in our case this is the quantity of messages) for *N* steps of the model evolution:

$$<\xi>_N = \frac{1}{N}\sum_{t=1}^{N}\xi(t)$$

Then the accumulated deviation of the measurement series *ξ(t)* from the average value $<\xi>_N$ is calculated:

$$X(t,N) = \sum_{u=1}^{t}(\xi(u) - <\xi>_N).$$

After that the difference between the maximal and the minimal accumulated deviation is calculated, which is called "range":

$$R(N) = \max_{1 \leq t \leq N} X(t,N) - \min_{1 \leq t \leq N} X(t,N).$$

Standard deviation is calculated according to the well-known formula:

$$S = \sqrt{\frac{1}{N}\sum_{t=1}^{N}(\xi(t) - <\xi>_N)^2}.$$

Experiment has shown that for many time series the following relation is true:
$$R/S = (N/2)^H$$

This is *H* coefficient that received the name of Hurst exponent, which is connected with the traditional "cellular" fractal dimension (*ρ*) by a simple relation:
$$\rho + H = 2.$$

The condition, at which Hurst exponent is connected with fractal "cellular" dimension in accordance with this formula, is the following: "*when the curve structure, describing a fractal function, is studied with high resolution, that is, in local limit*" [8]. Another important condition consists in function's self-affinity. For information streams this property is construed as self-similarity, arising as a result of processes of information streams' development. The time series, built on the grounds of powerful topic-based information streams, wholly satisfy this condition [10].

It is known, that Hurst exponent represents persistency degree – the process' inclination towards trends (as distinct from common Brownian movement). The value *H* > ½ means that the process' dynamics in the past, directed to the particular side, will most probably entail continuation of the movement in the same direction. If $H < ½$, it is forecasted that the process will change its direction. $H = ½$ means uncertainty.

Figure 5 shows the diagrams of *H* parameter change for the case-studies, considered above (*a* – case-study 1, *b* – case-study 2, *c* – case-study 3, *d* – case-study 4).

Change of the model parameters in case-studies 1 and 2 caused Hurst exponent to change from 0.73 to 0.85. So, we can come to a conclusion that an increase in the probability of messages "reposting" leads to higher persistency, that is, predictability of processes, related to the scope of evolution. A change in the model parameters leads to a spike in the dynamics of Hurst exponent, which, however, stabilizes, when the model parameters are stabilized.

In case of the time series, generated by the request "Tesla", the dynamics of Hurst parameter clearly indicates the time moment, when the information stream changes its behavior fundamentally. In this case the persistency level of the time-series is rather high (0.83), it is well-predicted and corresponds to model case-study 2.

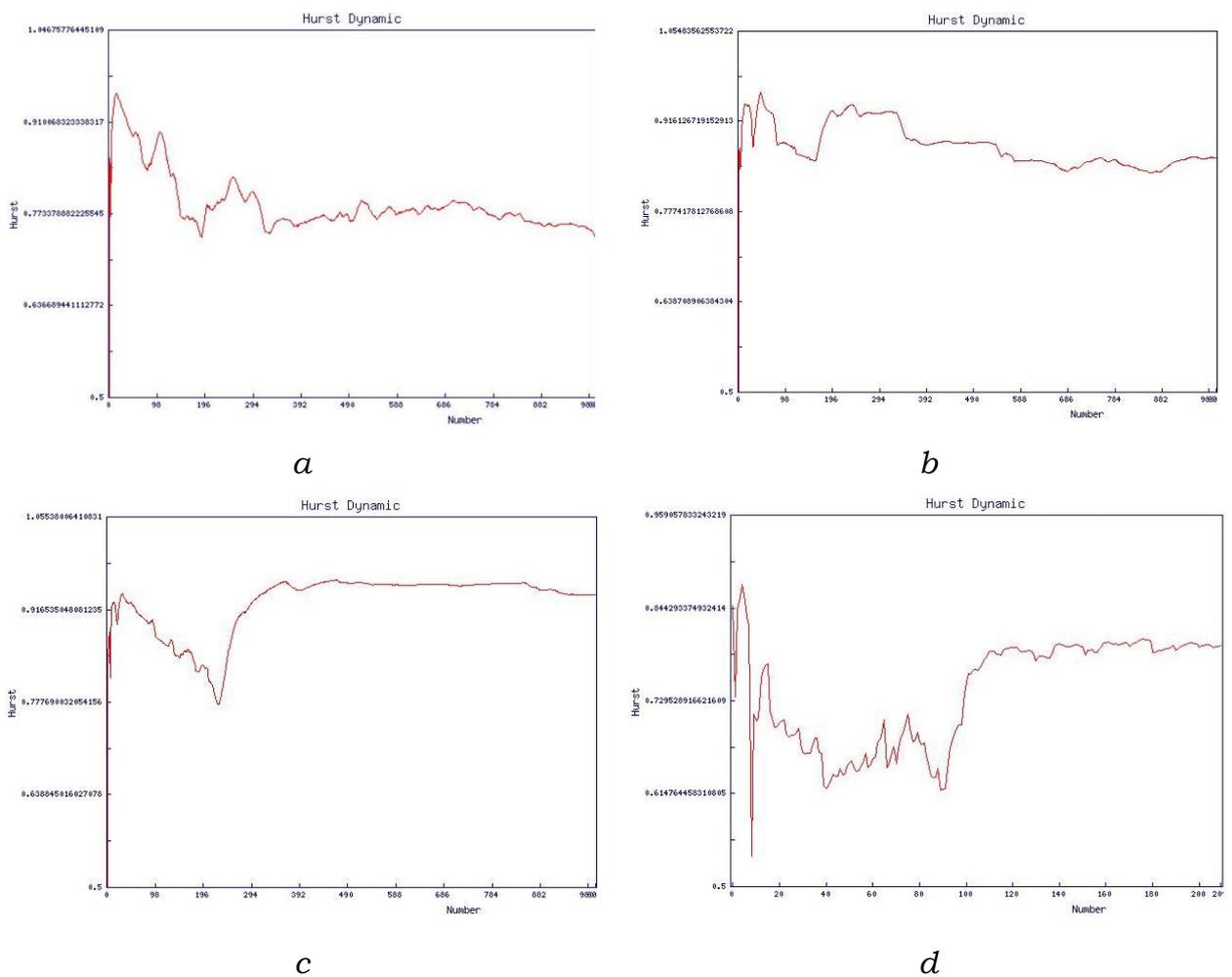

*a*          *b*

*c*          *d*

Fig. 5. Dynamics of Hurst exponent depending upon the evolution step

**Conclusions**

1. A model has been suggested, corresponding to the real information process. The authors have suggested an approach to modeling and further forecast of real information streams by way of changing the model parameters during its operation.
2. Along with affinity of general statistical characteristics (average value, standard deviation), similarity of wavelet-spectrograms, the validity of the multiagent model is confirmed by the coincidence of Hurst parameter values, with the probability of "likes" per agent equal to 0.05 and "reposts" from an agent – equal to 0.05.
3. With the help of the model and case-studies it has been shown, that it is possible to reveal changes in behavior of real information streams by analyzing changes in the dynamics of Hurst exponent.
4. The diagram of Hurst exponent dynamics has been compared with the wavelet-scaleogram. A more effective algorithm of Hurst exponent evaluation permits to recommend constant observation over this parameter dynamics in course of analytical work. Besides, it allows forecasting the information streams' behavior on the grounds of Hurst parameter value.